\documentclass[prd,aps,a4paper,eqsecnum,nofootinbib,floatfix,twocolumn]{revtex4}  

\newif\ifusesec
\usesectrue  
   
\usepackage{graphicx} 
\usepackage{amsmath,amsfonts,amssymb}

\newcommand{\beq}{\begin{equation}}
\newcommand{\eeq}{\end{equation}}
\newcommand{\bea}{\begin{eqnarray}}
\newcommand{\eea}{\end{eqnarray}}

\begin{document}

\title{Predicting today's cosmological constant via the Zel'dovich-Holographic connection}

\author{Pablo G. Tello$^1$, Donato Bini$^{2,3}$, Stuart Kauffman$^4$, Sauro Succi$^{2,5,6}$}
  \affiliation{
$^1$CERN, Geneva, Switzerland\\
$^2$Istituto per le Applicazioni del Calcolo \lq\lq M.~Picone,\rq\rq CNR, I-00185 Rome, Italy\\
$^3$INFN, Sezione di Roma Tre, I-00146 Rome, Italy\\
$^4$Institute for Systems Biology, Seattle, WA 98109, USA\\
$^5$
Istituto Italiano di Tecnologia, 00161 Rome, Italy\\
$^6$Physics Department, Harvard University, Cambridge,USA\\
}

\date{\today}

\begin{abstract}
This  Letter proposes a solution of the Vacuum Energy and 
the Cosmological Constant (CC)  paradox based on the Zel'dovich's ansatz, which states 
that the observable contribution to the vacuum energy density is given by the gravitational 
energy of virtual particle-antiparticle pairs, continually generated and annihilated in 
the vacuum state.  The novelty of this work is the use of an ultraviolet cut-off length based on 
the Holographic Principle, which is shown to yield current values of the CC in good 
agreement with experimental observations.
 \end{abstract}

\maketitle

\section{Introduction}

The Cosmological Constant (CC) problem or Vacuum Catastrophe stands for the stark mismatch 
between the currently observed values of the vacuum energy density 
(the small value of the CC) and theoretical large value of zero-point energy 
suggested by quantum field theory. 
It is also associated with a possible explanation for the dark energy driving 
the Universe accelerated expansion. Its theoretical value should therefore 
match observations. 

Unfortunately, due to about 120 orders of magnitude mismatch, it bears the 
reputation of \lq\lq the worst prediction in the history of 
physics\rq\rq~\cite{WOR}, see also~\cite{1,Bousso:2007gp,Weinberg:1988cp,Carroll:2000fy,Sola:2013gha}. 
This note sets out to calculate and explain the current experimental values of the CC by revisiting an 
original idea proposed by Zel'dovich and combining it with the Holographic Principle.

\section{The Cosmological Constant Paradox}

Despite being responsible for showing everyone that space-time is a dynamic entity, co-evolving with the matter that inhabits it, Einstein, for once, was not prepared for the idea that the entire Universe could be a dynamic entity as well. As a result, when faced with the irrefutable evidence that his equations did not admit a static universe as a solution, he resolved to add an \lq\lq ad-hoc" term, the CC, for obtaining one. Shortly later, however, it was for experimental data to show that our Universe is actually expanding, at which point he famously termed the CC his \lq\lq biggest blunder." However, to say with Joyce, \lq\lq errors are the portal of discovery," and the CC has taken central stage in modern physics, mostly because of its potential connections with dark energy and the accelerated expansion of the Universe. Not without a huge riddle, though: the CC has dimensions of an inverse length squared and since its physical origin is generally attributed to spacetime fluctuations at the Planck scale, it is natural to assume that its value in Planck units 
should be of order 1, namely:
\beq
\label{eq:1}
\Lambda L_{\rm P}^2\sim 1\,.
\eeq
 
By contrast,  for the product $\Lambda L_{\rm P}^2$  
cosmological observations deliver a value of $\sim 10^{-122}$, namely $122$ orders of magnitude smaller, making of \eqref{eq:1}, as mentioned before, \lq\lq the worst prediction ever in the history of physics\rq\rq \cite{1,Bousso:2007gp,Weinberg:1988cp,Carroll:2000fy,Sola:2013gha}. Despite intensive efforts, the puzzle is still standing.
Here, we begin by observing that $10^{-122}$ is surprisingly close to the square of the ratio of the Planck length and the Universe radius $10^{2(-35-27)}= 10^{-124}$, thereby providing a strong clue 
towards a theory where the \lq\lq natural" Eq. \eqref{eq:1} would 
be replaced by a much more accurate prediction:
\beq
\label{eq:2}
\Lambda L_{\rm P}^2=\left(\frac{L_{\rm P}}{L}\right)^2 \,,
\eeq
where $L\sim 10^{27}\, m$ is the current radius of the Universe.

In the following, it is shown that the above relation is precisely what one obtains 
by a straightforward combination of a previous argument by Zel'dovich,  with the Holographic Principle.

\section{Revisiting Zel'dovich's ansatz}

Zel'dovich argued that since the bare zero-point energy is unobservable, the observable contribution to the vacuum energy density, $e_v$, is given by the gravitational energy of virtual particle-antiparticle pairs, continually 
generated and annihilated in the vacuum state \cite{Zeldovich:1967gd,Zeldovich:1968ehl}. Therefore:
\beq
\label{eq:3}
e_v (r)\sim \frac{ Gm^2 (r)}{r } \frac{1}{r^3}\,.
\eeq
In the expression above, also according to Zel'dovich, the vacuum 
contains particles with an effective density $m(r)/r^3$. 
Additionally, by considering the Compton's expression for the wavelength, the effective 
mass of the particles at scale $r$ is taken as $m(r) \sim  \hbar/(cr)$. 
Substituting this in Eq. \eqref{eq:1}, and defining a local CC as:
\beq
\label{eq:4}
\Lambda(r)=\frac{ G e_v (r)}{c^4}\,, 
\eeq
one readily obtains:
\beq
\label{eq:5}
\Lambda L_{\rm P}^2\sim \left(\frac{ L_{\rm P} }{r }\right)^6\,,
\eeq
where $L_{\rm P}=(\hbar G/c^3)^{1/2}$ is the Planck length.
Next, we observe that the measured CC is likely to result from the average of the local CC
over the full spectrum of active scales \cite{Barrow:2011zp}, ranging from 
a UV cutoff to an IR one, which we shall be taken  
here as the {\it current} radius of the Universe. 
It is worth emphasizing that  in the present approach, such scales are not intended as 
regulatory devices to tame infinities but bear a physical meaning instead. 
They fix the boundaries of the spectrum  of dynamically active scales arising from 
the collective motion of the  nonlinearly interacting effective degrees of freedom\cite{NOI}.
As a mere analog,  in fluid turbulence the IR cutoff is the macroscopic scale $L$ of the
problem,  the molecular mean fee path $L_{\mu}$ is the underlying microscale, and the
Kolmogorov dissipative length  $L_d=L^{1/4}L_{\mu}^{3/4} $ represents  the shortest 
dynamically active scale supporting coherent hydrodynamic motion. 
The effective UV cutoff of turbulence is therefore provided by 
$L_d > L_{\mu}$ rather than  $L_{\mu}$, consistently with the macroscopic (supramolecular), 
nature of fluid turbulence as a self-interacting classical vector field theory \cite{FRISCH}.

Further to be noted,  the steep $1/r^6$ dependence 
(intriguingly the same exponent of the attractive branch of molecular Lennard-Jones 
interactions \cite{9}), implies that this average is largely dominated by the UV cutoff, namely:
\beq
\label{eq:6}
\Lambda L_{\rm P}^2\sim \left(\frac{ L_{\rm P} }{ L_{\rm UV}}\right)^6\,,
\eeq
where $L_{\rm UV}$ denotes the (yet unspecified) UV cutoff length. 
To fix the latter we resort to the Holographic Principle, which states that the 
minimum observable length scale is not the Planck length itself but 
a much larger holographic scale, given by \cite{Stephens:1993an,Susskind:1994vu,Ng:1993jb}:
\beq
\label{eq:6bis}
L_{\rm H}=L^{1/3} L_{\rm P}^{2/3}\,.
\eeq
Hence, we stipulate
\beq
\label{eq:6ter}
L_{\rm UV}=L_{\rm H} \,. 
\eeq
Inserting Eq. \eqref{eq:6ter} in Eq. \eqref{eq:5}, and taking into account 
the expression \eqref{eq:7}, we finally obtain:
\beq
\label{eq:7}
\Lambda L_{\rm P}^2=\frac{L_{\rm P}^6}{L^2 L_{\rm P}^4}=\left(\frac{L_{\rm P}}{L}\right)^2\,, 
\eeq
which is exactly the sought relation \eqref{eq:2}.
Such an expression shows that the current CCP (CC in Planck units) amounts to the second order 
term in a series of the cosmological smallness parameter $L_{\rm P}/L$. 
The fact that the second order term is entirely responsible for the value of the CCP 
reflects the $r^{-6}$ Zel'dovich decay in space, along with the $(2/3, 1/3)$ UV-IR 
exponents structure of the holographic scale, $L_{\rm H}$,  with no need 
of invoking any fine-tuning argument.

The present approach also provides a neat criterion to rule 
out other sources of vacuum energy.   
For instance,  the energy density of Casimir fluctuations is 
given by $e_{\rm Cas}(r) \sim \hbar c/r^4 $ \cite{Gambassi:2008beg,Kauffman:2021zlv}, yielding
$\Lambda L_{\rm P}^2 = (\frac{L_{\rm P}}{r})^4$,  namely, upon taking $r \sim L_{\rm UV}=L_{\rm H}$ and recalling Eq. \eqref{eq:6}, 
$\Lambda L_{\rm P}^2 = (\frac{L_{\rm P}}{L})^{4/3} \sim 10^{-83}$,  $40$ orders of magnitude too large.

In full generality,  the CCP associated with a vacuum energy scaling like $1/r^n$,  
is given by $(L_{\rm P}/L_{\rm UV})^n$,  which recovers the 
desired value $(L_{\rm P}/L)^2$ under the condition
\begin{equation}
\label{LUV}
L_{\rm UV} = L_{\rm P}^{1-2/n} L^{2/n}\,.
\end{equation}

This is clearly satisfied by the Zel'dovic-Holographic combination ($n=6$).

For the Casimir case, $n=4$, one computes $L_{\rm UV}=L_{\rm P}^{1/2} L^{1/2}$ ($L_{\rm UV}$, 
i.e.  the geometrical mean of $L_{\rm P}$ and $L$), meaning
that the cutoff should lie logarithmically midway between the Planck and the
Universe scale, i.e.,  $L_{\rm UV} \sim 10^{-4}$ meters.
We are not aware of any theoretical argument supporting such a UV cutoff 
based on Casimir physics.
Finally, the \lq\lq natural" choice $L_{\rm UV}=L_{\rm P}$,  corresponds 
to $n \to \infty$, an infinitely steep decay,  which does not look natural at all from the 
perspective discussed in this Letter.  

\section{Time Dependence}
All along this text, we have deliberately referred to the IR cut-off, $L$, as to 
the {\it current} radius  of our Universe,  in order to emphasize that
the present analysis does not encompass Universe's entire expansion chronology. 
In other words,  our explanation does not cover the value of the CC across
full time span since the Big Bang until now, but it only addresses the value of the
CC at the current time.
It does so, though, by proposing an alternative and possibly 
more economic explanation (in terms of assumptions) as compared
to previous ones  \cite{REF1,REF2}.

As it is well known, our Universe expansion chronology is parametrized 
by a dimensionless quantity,  known as the cosmic scale factor $a(t)$. 
Based on its time dependence, three characteristic eras can be distinguished: a 
radiation-dominated era encompassing the time scale from  
inflation until about 47,000 years after the Big Bang, where $a(t) \sim t^{1/2}$; 
a matter-dominated era, between about 47,000 years and 9.8 billion years after the 
Big Bang, where $a(t) \sim t^{2/3}$; and finally, the so called dark energy dominated 
era in which $a(t) \sim e^{H_0t}$,  ($H_0$ being the actual value of the Hubble "constant") and  
where our Universe is currently undergoing an accelerated expansion as 
suggested by observations \cite{REF3,REF4,REF5,REF6,REF7}. 
In the early universe, the mass-energy density effect was larger
 than the cosmological constant one, so the universal expansion was slowing 
 down (note that any power-law expansion implies a $1/t$ decay of the 
 Hubble parameter $H=\frac{\dot a}{a}$). 
 However, at around 6 billion years after the Big Bang, the mass-energy effect 
 became so diluted that the cosmological constant one took over. 
 As the universe evolved further, the mass-energy effect became less and 
 less important as compared to the cosmological constant effect, as confirmed by 
 experimental sources \cite{REF4}. 

Finally,  we note that the experimental evidence of a positive and small CC together 
with a potential eternal expansion of our Universe opens up the possibility that our 
Universe may asymptotically approach  a De Sitter one \cite{REF8}. 
Meaning a universe with no ordinary matter content but with a positive 
cosmological constant driving its expansion. 
In this context, our treatment might also offer a possible clue towards the 
explanation of the value of the CC in the mid-term and far-future regimes of our Universe.

\section{Conclusions}
In this Letter, we have proposed a straightforward solution of 
the vacuum energy and the CC paradox, based on the Zel'dovich's 
ansatz combined  with the Holographic Principle. 
The result is in nearly quantitative agreement with the experimental value, which is rather 
remarkable considering the simple nature of the supporting assumptions. 
This result suggests that, as originally proposed by Zel'dovich, the observable 
vacuum energy density today is given by the gravitational energy of virtual particle-antiparticle 
pairs, continually produced and annihilated in the vacuum state. 
Nevertheless, this argument alone does not suffice, as it requires a merger 
with the Holographic Principle, in order to select the appropriate UV cut-off length. 

It should also be pointed out that Zel'dovich needed to consider the proton mass 
as the \lq\lq typical" mass scale for producing a reasonably good order of magnitude result 
without a proper justification. Even then, his ansatz remained off the modern value by nine orders of magnitude. 
The approach suggested here does not necessitate any such restriction and provides 
a considerably better agreement with the experimental results.
Such a simplicity might represent a potential indication of 
its plausibility \lq\lq in the spirit of Occam's razor."

\section*{Acknowledgments}
DB thanks   ICRANet for partial support. 
DB also acknowledges sponsorship of the Italian Gruppo Nazionale 
per la Fisica Matematica (GNFM) of the Istituto Nazionale di Alta Matematica (INDAM).
SS kindly acknowledges funding from the European Research
Council under the Horizon 2020 Programme Grant Agreement n. 739964 (\lq\lq COPMAT").
All authors are grateful to Prof. David Spergel for very valuable remarks and criticism.


\begin{thebibliography}{00}

\bibitem{WOR}  
M.~P.~Hobson, G.~P.~Efstathiou and A.~N.~Lasenby,
General Relativity: An introduction for physicists, 
Cambridge University Press, 
(2006).

\bibitem{1} 
S.~E.~Rugha and H.~Zinkernagel, 
``The quantum vacuum and the cosmological constant problem," 
Studies in History and Philosophy of Modern Physics \textbf{33}, 663 (2002)  

\bibitem{Bousso:2007gp}
R.~Bousso, 
``TASI Lectures on the Cosmological Constant,''
Gen. Rel. Grav. \textbf{40}, 607-637 (2008)
[arXiv:0708.4231 [hep-th]].

\bibitem{Weinberg:1988cp}
S.~Weinberg,
``The Cosmological Constant Problem,''
Rev. Mod. Phys. \textbf{61}, 1-23 (1989)
 
\bibitem{Carroll:2000fy}
S.~M.~Carroll,
``The Cosmological constant,''
Living Rev. Rel. \textbf{4}, 1 (2001)
[arXiv:astro-ph/0004075 [astro-ph]].

\bibitem{Sola:2013gha}
J.~Sola,
``Cosmological constant and vacuum energy: old and new ideas,''
J. Phys. Conf. Ser. \textbf{453}, 012015 (2013)
[arXiv:1306.1527 [gr-qc]].

\bibitem{Zeldovich:1967gd}
Y.~B.~Zeldovich,
``Cosmological Constant and Elementary Particles,''
JETP Lett. \textbf{6}, 316 (1967)

\bibitem{Zeldovich:1968ehl}
Y.~B.~Zel'dovich, A.~Krasinski and Y.~B.~Zeldovich,
``The Cosmological constant and the theory of elementary particles,''
Sov. Phys. Usp. \textbf{11}, 381-393 (1968)

\bibitem{Barrow:2011zp}
J.~D.~Barrow and D.~J.~Shaw,
``The Value of the Cosmological Constant,''
Gen. Rel. Grav. \textbf{43}, 2555-2560 (2011)
[arXiv:1105.3105 [gr-qc]].

\bibitem{NOI}
D. Bini,  S. Kauffman, S. Succi and P. Tello,
``First Post-Minkowskian approach to Turbulent Gravity'',
[arXiv:2208.03572 [gr-qc]] 

\bibitem{FRISCH}
U. Frisch,  
``Turbulence: the Legacy of A.N. Kolmogorov,''
Cambridge University  Press (1996)

\bibitem{9} B. Smit and D. Frenkel, 
{\it Understanding Molecular Simulation: from algorithms to applications}, 
Academic Press, NY, 2$^{nd}$ Edition, 2001.

\bibitem{Stephens:1993an}
C.~R.~Stephens, G.~'t Hooft and B.~F.~Whiting,
``Black hole evaporation without information loss,''
Class. Quant. Grav. \textbf{11}, 621-648 (1994)
[arXiv:gr-qc/9310006 [gr-qc]].

\bibitem{Susskind:1994vu}
L.~Susskind,
``The World as a hologram,''
J. Math. Phys. \textbf{36}, 6377-6396 (1995)
[arXiv:hep-th/9409089 [hep-th]].

\bibitem{Ng:1993jb}
Y.~J.~Ng and H.~Van Dam,
``Limit to space-time measurement,''
Mod. Phys. Lett. A \textbf{9}, 335-340 (1994)

\bibitem{Gambassi:2008beg}
A.~Gambassi,
``The Casimir effect: From quantum to critical fluctuations,''
J. Phys. Conf. Ser. \textbf{161}, 012037 (2009)
[arXiv:0812.0935 [cond-mat.stat-mech]].

\bibitem{Kauffman:2021zlv}
S.~Kauffman, S.~Succi, A.~Tiribocchi and P.~G.~Tello,
``Playing with Casimir in the vacuum sandbox,''
Eur. Phys. J. C \textbf{81}, no.10, 941 (2021)
[arXiv:2102.11326 [quant-ph]].

\bibitem{REF1}
P.~C.~W. Davies and S.~D.~Unwin, 
``Why is the cosmological constant so small?," 
Proc. R. Soc. Lond. A  {\bf 377}, n. 1769, 147-149 (1981) 

\bibitem{REF2}
P.~J.~Steinhardt and N.~Turok,
``Why the cosmological constant is small and positive,''
Science \textbf{312}, 1180-1182 (2006)
[arXiv:astro-ph/0605173 [astro-ph]].

\bibitem{REF3}
Planck Collaboration (2020),
Planck 2018 results. VI. Cosmological parameters. 
Astronomy \& Astrophysics. 641: A6. 

\bibitem{REF4}
A.~G.~Riess et al., 
``The Farthest Known Supernova: Support for an Accelerating Universe and 
a Glimpse of the Epoch of Deceleration," 
The Astrophys. J., \textbf{560},  49-71 (2001)  

\bibitem{REF5} 
 S.~Perlmutter et al., 
``Measurements of $\Omega$ and $\Lambda$ from 42 High-Redshift Supernovae," 
 Astrophys. J. \textbf{517}, 565-586   (1999)

\bibitem{REF6} 
D.~N.~Spergel et al., 
``First-Year Wilkinson Microwave Anisotropy Probe (WMAP) Observations: Determination of Cosmological Parameters," 
Astrophys. J. Suppl. Ser. \textbf{148}, 175-194, (2003) 

\bibitem{REF7} 
D.~N. Spergel et al., 
``Three-Year Wilkinson Microwave Anisotropy Probe (WMAP) Observations: Implications for Cosmology," 
Astrophys. J. Supplement Series, \textbf{170}, 377-408  (2007)

\bibitem{REF8}
R.~Bousso, 
``The Cosmological Constant Problem, Dark Energy, and the Landscape of String Theory," 
Proceedings of the Subnuclear Physics: 
Past, Present and Future, Pontificial Academy of Sciences: Vatican City, Vatican,  2011.


\end{thebibliography}
\end{document}